\documentclass[a4paper]{jpconf}
\usepackage{graphicx}
\usepackage{epstopdf}
\usepackage{subfigure}
\usepackage{textcomp}

\def\bkR{{\rm I\kern-.17em R}}
\def\bkC{{\rm \kern.24em \vrule width.05em height1.4ex depth-.05ex \kern-.26em C}}
\def\to{\rightarrow}

\def\be{\beta}

\def\frac#1#2{{\textstyle{{#1}\over {#2}}}}
\def\laq{\raise 0.4 ex \hbox{$<$}\kern -0.8 em\lower 0.62 ex\hbox{$\sim$}}
\def\gaq{\raise 0.4 ex \hbox{$>$}\kern -0.7 em\lower 0.62 ex\hbox{$\sim$}}

\def\be{\begin{equation}}
\def\ee{\end{equation}}
\def\ba{\begin{eqnarray}}
\def\ea{\end{eqnarray}}

\def\dalemb#1#2{{\vbox{\hrule height.#2pt
        \hbox{\vrule width.#2pt height#1pt \kern#1pt \vrule width.#2pt}
        \hrule height.#2pt}}}

\def\dalemb#1#2{{\vbox{\hrule height.#2pt
        \hbox{\vrule width.#2pt height#1pt \kern#1pt \vrule width.#2pt}
        \hrule height.#2pt}}}

\def\gtorder{\mathrel{\raise.3ex\hbox{$>$}\mkern-14mu
             \lower0.6ex\hbox{$\sim$}}}
\def\ltorder{\mathrel{\raise.3ex\hbox{$<$}\mkern-14mu
             \lower0.6ex\hbox{$\sim$}}}

\begin{document}

\title{\bf Noncommutative Black Holes and the Singularity Problem \footnote{Based on a talk presented by CB at ERE 2010, Granada, Spain, 6th-10th September 2010.}}

\author{C Bastos$^{1}$, O Bertolami$^{1,2}$, N C Dias$^{3,4}$ and J N Prata$^{3,4} $}

\address{$^1$ Instituto de Plasmas e Fus\~ao Nuclear, Instituto Superior T\'ecnico, Avenida Rovisco Pais 1, 1049-001 Lisboa, Portugal \\   
  $^2$ Departamento de F\'isica e Astronomia, Faculdade de Ci\^ encias da Universidade do Porto, Rua do Campo Alegre 687, 4169-007 Porto, Portugal \\
  $^3$ Departamento de Matem\'atica, Universidade Lus\'ofona de Humanidades e Tecnologias, Avenida Campo Grande, 376, 1749-024 Lisboa, Portugal \\
  $^4$ Grupo de F\'isica Matem\'atica, Universidade de Lisboa, Avenida Prof. Gama Pinto 2, 1649-003, Lisboa, Portugal}

\ead {cbastos@fisica.ist.utl.pt, orfeu.bertolami@fc.up.pt, ncdias@mail.telepac.pt, joao.prata@mail.telepac.pt}

\begin{abstract}
{A phase-space noncommutativity in the context of a Kantowski-Sachs cosmological model is considered to study the interior of a Schwarzschild black hole. Due to the divergence of the probability of finding the black hole at the singularity from a canonical noncommutativity, one considers a non-canonical noncommutativity. It is shown that this more involved type of noncommutativity removes the problem of the singularity in a Schwarzschild black hole. } 
\end{abstract}

\section{Introduction}

It is expected that a quantum theory of gravity may resolve the black hole (BH) singularity problem. However, as a quantum theory of gravity is beyond to reach, a quantum cosmology approach based on a minisuperspace approximation might provide some insight into problems of this nature. Assuming that the ultimate structure of space-time should be determined by quantum gravity and that should be noncommutative (NC), then noncommutativity in a quantum cosmological level should be considered. In fact, noncommutativity of space-time is shown to be relevant in the context of BHs physics \cite{Bastos5}. One of the most interesting features of this noncommutativity is that it allows for a square integrable wave function, solution of the noncommutative Wheeler-DeWitt (WDW) equation. 

In this contribution one presents a phase-space non-canonical NC extension of the Kantowski-Sachs (KS) cosmological model \cite{Bastos2} and its application to study the singularity of a Schwarzschild BH \cite{Bastos5}. In the interior of a Schwarzschild BH, for $r<2M$, time and radial coordinates are interchanged ($r\leftrightarrow t$) and space-time is described by the metric:
\be\label{eq0.2}
ds^2=-\left({2M\over t}-1\right)^{-1}dt^2+\left({2M\over t}-1\right)dr^2+t^2(d\theta^2+\sin^2\theta d\varphi^2)~.
\ee
That is, an isotropic metric turns into an anisotropic one, implying that the interior of a Schwarzschild BH can be described by an anisotropic cosmological space-time. Thus, the metric (\ref{eq0.2}) can be mapped into the KS cosmological model \cite{AK}, which, in the Misner parametrization, can be written as
\be\label{eq1.4}
ds^2=-N^2dt^2+e^{2\sqrt{3}\beta}dr^2+e^{-2\sqrt{3}(\beta+\Omega)}(d\theta^2+\sin^2{\theta}d\varphi^2)~,
\ee
where $\Omega$ and $\beta$ are scale factors, and $N$ is the lapse function. The following identification for $t<2M$,
\be\label{eq0.3}
N^2=\left({2M\over t}-1\right)^{-1} \quad, \quad
e^{2\sqrt{3}\beta} = \left({2M\over t}-1\right) \quad , \quad
e^{-2\sqrt{3}\beta}e^{-2\sqrt{3}\Omega}=t^2~,
\ee
allows for mapping the metric Eq. (\ref{eq1.4}) into the metric Eq. (\ref{eq0.2}).

One considers that the two scale factors $\Omega$ and $\beta$, and the corresponding canonical conjugated momenta $P_{\Omega}$ and $P_{\beta}$ do not commute, and
\ba\label{eq1.5}
\left[\hat{\Omega}, \hat{\beta} \right]\! &=&\! i \theta \left( 1 + \epsilon \theta \hat{\Omega} + {\epsilon \theta^2\over{1 + \sqrt{1- \xi}}} \hat{P}_{\beta} \right)\\
\left[\hat{P}_{\Omega}, \hat{P}_{\beta} \right]\! &=&\! i \left( \eta  + \epsilon (1 + \sqrt{1 - \xi})^2 \hat{\Omega} + \epsilon \theta (1 + \sqrt{1- \xi}) \hat{P}_{\beta} \right)\nonumber\\
\left[\hat{\Omega}, \hat{P}_{\Omega} \right]\! &=&\! \left[\hat{\beta}, \hat{P}_{\beta} \right]\! =\!i  \left( 1 + \epsilon \theta (1 + \sqrt{1-
\xi}) \hat{\Omega} + \epsilon \theta^2 \hat{P}_{\beta} \right)\nonumber,
\ea
where $\theta$, $\eta$ and $\epsilon$ are positive constants and $\xi = \theta \eta <1$, a quite physical condition \cite{Bastos7}. The noncommutative WDW (NCWDW) equation of the KS cosmological model obtained in Ref. \cite{Bastos6} is considered to describe the interior of the BH. The remaining commutation relations vanish. For $\epsilon\neq0$ it implies that the noncommutative commutation and uncertainty relations are themselves position and momentum dependent. The case $\epsilon=0$ corresponds to the canonical phase-space noncommutativity \cite{Bastos1,Bastos2,Bastos5}. One can relate this NC algebra with the Heisenberg-Weyl algebra, via a suitable map. Suppose that $\left(\hat{\Omega}_c, \hat{P}_{\Omega_c}, \hat{\beta_c}, \hat{P}_{\beta_c} \right)$ obey the HW algebra, $\left[\hat{\Omega}_c, \hat{P}_{\Omega_c} \right] = \left[\hat{\beta}_c, \hat{P}_{\beta_c} \right] =i$. Then a suitable transformation would be:
\ba\label{eq1.7}
\hat{\Omega} \!\!&=&\!\! \lambda \hat{\Omega}_c -  {\theta\over2\lambda} \hat{P}_{\beta_c} + E \hat{\Omega}_c^2\hspace{0.2cm},\hspace{0.2cm}\hat{\beta} = \lambda \hat{\beta}_c + {\theta\over2 \lambda} \hat{P}_{\Omega_c} \nonumber\\
\hat{P}_{\Omega}\!\! &=&\!\! \mu \hat{P}_{\Omega_c} +  {\eta\over2\mu} \hat{\beta}_c \hspace{0.2cm},\hspace{0.2cm}\hat{P}_{\beta} = \mu \hat{P}_{\beta_c} -  {\eta\over2 \mu} \hat{\Omega}_c + F \hat{\Omega}_c^2.
\ea
Here, $\mu$, $ \lambda$ are real parameters such that $(\lambda \mu)^2 - \lambda \mu + \frac{\xi}{4} =0\Leftrightarrow 2 \lambda \mu = 1 \pm \sqrt{1- \xi}$, and one chooses the positive solution given the invariance of the physics on the D map \cite{Bastos1}, and
\be\label{eq1.9}
E = - {\theta\over{1 + \sqrt{1- \xi}}} F\;,\;F = - {\lambda\over\mu} \epsilon \sqrt{1- \xi} \left(1 + \sqrt{1- \xi} \right)~.
\ee
Of course, the inverse transformation can be obtained \cite{Bastos6}.

As in the canonical NC KS cosmological model \cite{Bastos2,Bastos5}, one obtains the NCWDW equation considering the canonical quantization of the KS Hamiltonian, the operator $\hat{A}$ that commutes with the Hamiltonian and the D map Eqs. (\ref{eq1.7}), \cite{Bastos6}
\ba\label{eq1.19}
&&\mu^2 {\cal R}''- 48 \exp \left( - {2 \sqrt{3}\over\mu} \Omega_c - 2 \sqrt{3} E \Omega_c^2 + {\sqrt{3} \theta a\over\mu \lambda} \right) {\cal R}- {2 \eta\over\mu} \left(\Omega_c+F\Omega_c^3\right){\cal R} + \nonumber\\&&+F^2 \Omega_c^4{\cal R} + a^2{\cal R} + \left({\eta^2\over\mu^2}+2 a F \right) \Omega_c^2{\cal R}=0~.
\ea
The dependence on $\beta_c$ has completely disappeared and one is left with an ordinary differential equation for
${\cal R} \left(\Omega_c \right)$. Through the substitution, $\Omega_c = \mu z$, ${d^2\over d \Omega_c^2} = {1\over\mu^2} {d^2\over d z^2}$ and ${\cal R} \left( \Omega_c (z) \right) := \phi_a (z)$
one obtains a second order linear differential equation, a Schr\"odinger-like equation:
\be\label{eq1.21}
- \phi_a'' (z) +V(z) \phi_a (z)= 0~,
\ee
where the potential function, V(z), reads:
\be\label{eq1.22}
V(z) = - \left( \eta z -a \right)^2 - F^2 \mu^4 z^4 - 2  F \mu^2(\eta z-a) z^2 + 48 \exp \left( -2 \sqrt{3} z -2 \sqrt{3} \mu^2 E z^2 + {\sqrt{3} \theta a \over\mu \lambda} \right).
\ee
Equation (\ref{eq1.21}) depends explicitly on the noncommutative parameters $\theta$, $\eta$, $\epsilon$ and the eigenvalue $a$. Notice that $E>0$. 

In this contribution one analyzes the singularity $t=0$ in the Schwarzschild BH with this NCWDW equation, Eqs. (\ref{eq1.21}) and (\ref{eq1.22}). In fact, with the asymptotic solutions of this NCWDW equation one is able to study the singularity of the BH. In the canonical NC approach, these type of solutions vanish in the neighbourhood of $t=0$, but the probability of finding the system at the singularity diverges \cite{Bastos5}. However, using the non-canonical noncommutativity, one actually can remove the singularity of the Schwarzschild BH \cite{Bastos6}.

\section{Singularity}

To study the singularity of the BH, one needs to examine the asymptotic limit of the NCWDW. From Eq. (\ref{eq1.22}) one finds that asymptotically $(z \to \infty)$ the potential function is dominated by the term
\be\label{eq1.24}
V(z) \sim - F^2 \mu^4 z^4,
\ee
which leads to $L^2(\bkR)$ solutions of the Schr\"odinger equation (\ref{eq1.21}). In fact, the Hamiltonian $H= - \frac{\partial^2}{\partial z^2} - F^2 \mu^4 z^4$ has a continuous and real spectrum \cite{Gitman} in $L^2$. The asymptotic form of the eigenfunction corresponding to the eigenvalue zero and eigenvalue $a$ of $\hat A$ is
 \be\label{eq1.25}
 \phi_a (z) \sim {1\over z} \exp \left[ \pm i \frac{F \mu^2}{3} z^3 \right]~.
 \ee
Thus, one can write a general solution of the NCWDW equation as
 \be\label{eq1.26}
 \psi (\Omega_c, \beta_c) = \int C(a) \psi_a (\Omega_c, \beta_c ) da~,
 \ee
 where $C(a)$ are arbitrary complex constants and $\psi_a (\Omega_c, \beta_c )$ is of the form:
 \be\label{eq1.27}
 \psi_a (\Omega_c, \beta_c ) = B \phi_a \left(\frac{\Omega_c}{\mu} \right) \exp \left[\frac{i \beta_c}{\mu}
 \left(a - \frac{\eta}{2 \mu} \Omega_c \right) \right]~.
 \ee
 Here $\phi_a(\Omega/\mu)$ is a the solution of Eq. (\ref{eq1.21}) and $B$ is a normalization constant such that:
 \be\label{eq1.27a}
 \int |\psi_a (\Omega_c, \beta_c )|^2  d\Omega_c = B^2 \int  |\phi_a \left(\frac{\Omega_c}{\mu} \right)|^2 d \Omega_c=1
 \ee
One fixes a constant $\beta_c$-hypersurface, given that the wave function is oscillatory in $\beta_c$. Thus, it is natural to consider a measure $d \xi=\delta (\beta- \beta_c) d \beta d\Omega_c$ in order to define the inner product in $L^2(\bkR^2)$ and the computation of probabilities. For the wave function Eq. (\ref{eq1.26}) one then finds:
 \be\label{eq1.27b}
 ||\psi||_{L^2(\bkR^2,d\xi)}\!\!=\!\!\!\left(\int \psi (\Omega_c, \beta ) \psi^* (\Omega_c, \beta ) d\xi \right)^{1/2}\!\!\!\!\!\!\! = \!\!\left( \int C(a)C^*(a')\!\! \left( \int \psi_a (\Omega_c, \beta_c ) \psi_{a'}^* (\Omega_c, \beta_c ) d\Omega_c \right)\!\! da da' \right)^{1/2}.
 \ee
The wave functions $\psi_a(\Omega_c,\beta_c)$ are solutions of a hyperbolic-type equation, and in general, they are not orthogonal to each other. So, one uses the Cauchy-Schwartz inequality to compute the $L^2(\bkR^2,d\xi)$-inner product between $\psi_a (\Omega_c, \beta_c )$ and $\psi_{a'}^* (\Omega_c, \beta_c )$. It follows that:
 \be\label{eq1.27c}
 ||\psi||_{L^2(\bkR^2,d\xi)} \le \int |C(a)| \, ||\psi_a (\Omega_c, \beta_c )||_{L^2(\bkR^2,d\xi)} da = \int |C(a)| da
 \ee
where, in the last step, the relation (\ref{eq1.27a}) was used. Thus, one finally concludes that for $C(a) \in L^1(\bkR)$ the wave function Eq. (\ref{eq1.26}) is squared integrable on constant $\beta_c$-hypersurfaces and the BH probability $P(r=0,t=0)$ at the singularity can now be calculated:
 \be\label{eq1.28}
P(r=0,t=0)= \lim_{\tilde\Omega_c, \beta_c \to + \infty} \int_{\tilde\Omega_c}^{+ \infty} \int_{-\infty}^{+\infty} | \psi({\Omega}_c, \beta)|^2 d\xi =\lim_{\tilde\Omega_c , \beta_c \to + \infty} \int_{\tilde\Omega_c}^{+ \infty} \left| \psi(\Omega_c,\beta_c) \right|^2 d{\Omega}_c =0~.
 \ee
Notice that in the last step, it was used the fact that $\psi(\Omega_c,\beta_c) \in L^2(\bkR^2,d\xi)$. Consequently, the integral in $\Omega_c$ vanishes, independently of the value of $\beta_c$.

\section{Conclusions}

In this contribution, a non-canonical phase-space NC extension of the KS cosmological model is used to study the singularity of a Schwarzschild BH. In contrast with the canonical NC studied in Ref. \cite{Bastos5}, this type of noncommutativity allows us to obtain square integrable solutions for the asymptotic NCWDW equation. In fact, this is the major property of the model. That is, the general solution of our NCWDW equation is square integrable on the constant $\beta_c$-hypersurfaces, and so one can use the same rules as in quantum mechanics to evaluate probabilities. In this case, one computes the probability of finding the BH at the singularity which is found to vanish.

\ack

\noindent The work of CB is supported by Funda\c{c}\~{a}o para a Ci\^{e}ncia e a Tecnologia (FCT) under the grant SFRH/BPD/62861/2009. The work of OB is partially supported by the FCT grant PTDC/FIS/111362/2009. NCD and JNP were partially supported by the grants PTDC/MAT/69635/2006 and PTDC/MAT/099880/2008 of FCT.

\end{document}